\documentclass[%
 reprint,
superscriptaddress,
preprintnumbers,
nofootinbib,
 amsmath,amssymb, 
 aps,
 prd,
 longbibliography,
]{revtex4-1}

\usepackage{cancel}
\usepackage{accents}
\usepackage{mciteplus,slashed}
\usepackage{amssymb,cancel,amsmath}
\usepackage{dcolumn}
\usepackage{bm}
\usepackage[caption=false]{subfig}
\usepackage{appendix}
\usepackage{physics,mathtools}
\usepackage{booktabs}
\usepackage{feynmp-auto}
\usepackage[T1]{fontenc}	
\usepackage{csvsimple}
\usepackage[hidelinks]{hyperref}
\usepackage[section]{placeins}
\usepackage[capitalise]{cleveref}
\usepackage{booktabs}
\usepackage{graphicx}
\usepackage{mathrsfs}
\graphicspath{{Figures/}}

\usepackage{float} 
\usepackage{comment}

\AtBeginDocument{
\heavyrulewidth=.08em
\lightrulewidth=.05em
\cmidrulewidth=.03em
\belowrulesep=.65ex
\belowbottomsep=0pt
\aboverulesep=.4ex
\abovetopsep=0pt
\cmidrulesep=\doublerulesep
\cmidrulekern=.5em
\defaultaddspace=.5em
}
\usepackage[dvipsnames]{xcolor}
\usepackage[normalem]{ulem}
\usepackage{fontawesome} 

\hypersetup{
    pdfnewwindow=true,      
    colorlinks=true,       
    linkcolor=blue,          
    citecolor=blue,        
    filecolor=blue,      
    urlcolor=blue        
}

\newcommand{\github}[1]{\href{https://github.com/#1}{\includegraphics[width=10pt]{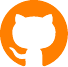}}}

\newcommand{\orcid}[1]{\begingroup
  \hypersetup{hidelinks}\href{https://orcid.org/#1}{\includegraphics[width=10pt]{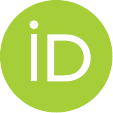}} \endgroup}

\begin{document}

\title{Final state radiation from high and ultrahigh energy neutrino interactions}

\author{Ryan Plestid \orcid{0000-0003-0779-7289}\,}
\email{rplestid@caltech.edu}
\affiliation{Walter Burke Institute for Theoretical Physics, California Institute of Technology, Pasadena, California 91125, USA}

\author{Bei Zhou \orcid{0000-0003-1600-8835}\,}
\email{beizhou@fnal.gov}
\affiliation{Theoretical Physics Department, Fermi National Accelerator Laboratory, Batavia, Illinois 60510, USA}
\affiliation{Kavli Institute for Cosmological Physics, University of Chicago, Chicago, Illinois 60637, USA}

\preprint{CALT-TH-2024-008}
\preprint{FERMILAB-PUB-24-0087-T}

\begin{abstract}
Charged leptons produced by high-energy and ultrahigh-energy neutrinos have a substantial probability of emitting prompt internal bremsstrahlung $\nu_\ell + N \rightarrow \ell + X + \gamma$. This can have important consequences for neutrino detection.  We discuss observable consequences at high- and ultrahigh-energy neutrino telescopes and the Large Hadron Collider's Forward Physics Facility. Logarithmic enhancements can be substantial (e.g., $\sim 20\%$) when either the charged lepton's energy or the rest of the cascade is measured. We comment on final state radiation's impacts on measuring the inelasticity distribution, $\nu/\bar{\nu}$ flux ratio, throughgoing muons, and double-bang signatures for high-energy neutrino observation. Furthermore, for ultrahigh-energy neutrino observation, we find that final state radiation increases the overall detectable energy by as much as 20\%, affects flavor measurements, and decreases the energy of both Earth-emergent tau leptons and regenerated tau neutrinos. Many of these have significant impacts on measuring neutrino fluxes and spectra. Finally, for the Large Hadron Collider's Forward Physics Facility, we find that final state radiation will impact future extractions of strange quark parton distribution functions. Final state radiation should be included in future analyses at neutrino telescopes and the Forward Physics Facility. 
 \end{abstract}

\maketitle

\section{Introduction}
 
High-energy (HE, i.e., $\sim$ 100~GeV--100~PeV) and ultrahigh-energy (UHE, i.e., $\gtrsim 100~{\rm PeV}$) neutrinos are important for both neutrino astrophysics and multimessenger astronomy~\cite{Ackermann:2022rqc, Ackermann:2019ows, LIGOScientific:2017ync, IceCube:2018dnn, IceCube:2022der, IceCube:2023ame}. They provide a unique window into extreme astrophysical environments.
They also probe neutrino interactions at high center-of-mass energies, and offer valuable tests of the Standard Model of particle physics~\cite{Klein:2019nbu, Ackermann:2022rqc, Reno:2023sdm}. For example, measurements of neutrino-nucleon scattering at TeV energies and above probe quantum chromodynamics (QCD) at small $x$~\cite{Anchordoqui:2006ta, Bertone:2018dse, Amoroso:2022eow, Cruz-Martinez:2023sdv} and offer unique opportunities to constrain strange-quark distribution functions inside the nucleon~\cite{NOMAD:2013hbk, DeLellis:2004ovi, Hou:2019efy, Faura:2020oom, Zhou:2021xuh}. Measurements of neutrino interactions at high energies can be obtained using naturally occurring neutrino fluxes with neutrino telescopes~\cite{IceCube:2017roe, Bustamante:2017xuy, IceCube:2020rnc, Valera:2022ylt, Esteban:2022uuw, Valera:2023ayh}. Similar measurements can also be performed at the Large Hadron Collider's (LHC's) Forward Physics Facility (FPF), which is exposed to neutrinos with energies as high as a few TeV~\cite{Anchordoqui:2021ghd, Feng:2022inv}. Furthermore, measurements of high- and ultrahigh-energy neutrinos, (including energy spectra~\cite{Anchordoqui:2005ey, Hooper:2007jr, Ng:2014pca, Ioka:2014kca, Bustamante:2020mep, Creque-Sarbinowski:2020qhz, Esteban:2021tub}, flavor ratios \cite{Mehta:2011qb, Arguelles:2015dca, Bustamante:2015waa, Shoemaker:2015qul, Ahlers:2020miq}, and arrival directionality~\cite{Zentner:2009is, Bai:2013nga, Reno:2021cdh} and timing \cite{Jacob:2006gn, Huang:2019etr, Addazi:2021xuf})  offer new probes of neutrino properties in a largely untested energy range and offer discovery opportunities for physics beyond the Standard Model~\cite{Ackermann:2019ows}.

\begin{figure}[t]
    \centering
    \includegraphics[width=\linewidth]{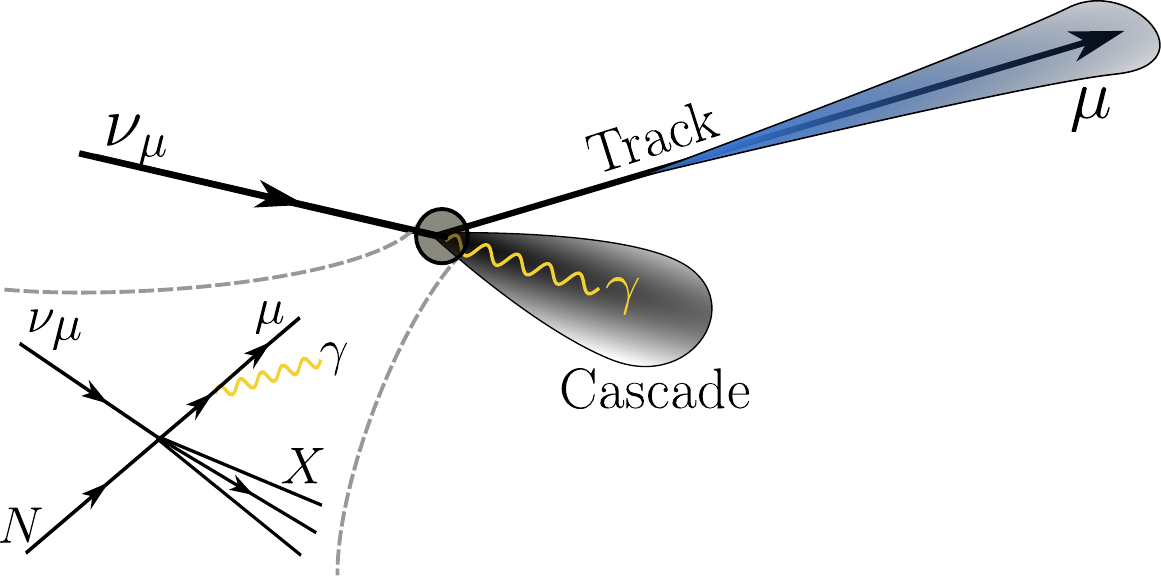}
    \caption{
    Effect of final state radiation on HE neutrino scattering (CCDIS+FSR) and detection, taking $\nu_\mu$ as an example.  A $\nu_\mu$ produces a high-energy muon, which emits a photon during CCDIS with a nucleon (bottom left). The photon is absorbed in the cascade, while the muon is observed as a track. In general, any measurements that can separate the charged lepton and cascade are subject to logarithmically enhanced QED radiative corrections. Tau neutrino detection is also significantly impacted by FSR, as discussed below.\label{cartoon} }
\end{figure}

For both astrophysics and particle physics applications, a precise description of neutrino interaction and detection plays a central role.
Statistical samples have grown, and will grow, with more and more HE and UHE neutrino telescopes being built or proposed (e.g., Refs.~\cite{KM3Net:2016zxf, GRAND:2018iaj, IceCube-Gen2:2020qha, P-ONE:2020ljt, PUEO:2020bnn, RNO-G:2020rmc, POEMMA:2020ykm, Romero-Wolf:2020pzh, Ye:2022vbk}; see Ref.~\cite{Ackermann:2022rqc} for a comprehensive overview) and the prospect of new data from the LHC's FPF~\cite{Feng:2022inv}. Consequently, precision goals are becoming increasingly stringent. This has motivated the study of subleading QCD corrections to neutrino-nucleus cross sections, with quoted uncertainties as small as $\sim 1\%$~\cite{Cooper-Sarkar:2011jtt, Connolly:2011vc, Xie:2023suk}. Other work has focused on subdominant interaction channels, such as $W$-boson and trident production~\cite{Seckel:1997kk, Alikhanov:2015kla, Gauld:2019pgt, Zhou:2019vxt, Zhou:2019frk, Zhou:2021xuh, Xie:2023qbn}.

In this work, we consider subleading corrections to charged-current deep-inelastic scattering (CCDIS) stemming from the emission of hard on-shell photons, i.e., final state radiation (FSR). This has been previously discussed in the context of the inelasticity distribution for accelerator \cite{DeRujula:1979grv} and astrophysical \cite{Sigl:1997cq} neutrinos.
We revisit FSR's impact on the inelasticity distribution and study many other phenomenological impacts for HE and UHE neutrino observation for the first time. 
As we discuss below, for neutrino energies above 100~GeV, quantum electrodynamic (QED) radiative corrections can be sizeable when observables are sensitive to FSR~\cite{Sigl:1997cq}, i.e., when the charged lepton ($\ell$) emits an energetic photon ($\gamma$) on top of the CCDIS, $\nu_\ell + N \rightarrow \ell + X + \gamma$. Here, we label the nucleon by $N$ and final state hadrons by $X$, and the relevant cross section is inclusive with respect to final state hadrons.

In \cref{cartoon}, we show a schematic of $\nu_\mu$ scattering (CCDIS+FSR) and detection as an example. (FSR also significantly impacts $\nu_\tau$ scattering and detection, as discussed below.)
High-energy muons travel macroscopic distances in matter (e.g., tens of km), whereas photons produce electromagnetic cascades over a distance scale set by the radiation length, $X_0 \sim 10~{\rm cm}$, similar to the hadronic cascades. In neutrino telescopes, muons are ``stripped bare'' of any prompt radiation, because the photon is absorbed in a local cascade, whereas the muon itself leaves a long track. In contrast to jet-like observables, the experimentally observed muon energy does not contain the prompt photon.

For many detection channels of HE and UHE neutrinos, the charged lepton and the primary (hadronic) cascade are separable. In these scenarios, relevant observables are not inclusive with respect to FSR, which is enhanced by Sudakov double logarithms. This is especially true for observables with large statistical samples that are binned as a function of energy, e.g., inelasticity distributions \cite{IceCube:2018pgc}. Therefore, radiative corrections must be carefully considered when discussing non-inclusive observables for sub-TeV to EeV neutrino detection. 
For a quick estimate, consider the Sudakov factor\footnote{There is a factor of two difference in the argument of the exponential relative to the classic Sudakov factor since there is one lepton, rather than two, in the final state.} (see, e.g., Ref.~\cite{Peskin:1995ev})
\begin{equation}
    \label{eq_sudakov}
    F_{S}(s,E_{\rm min}) \sim \exp\qty[-\frac{\alpha}{2\pi} \log(\frac{s}{m_\ell^2}) \log(\frac{E_\ell^2}{E_{\rm min}^2}) ] ~, 
\end{equation}
which gives the probability to {\it not} radiate any photons above $E_{\rm min}$ in a collision with center-of-mass energy $\sqrt{s}$ and final-state charged-lepton energy $E_\ell$. Taking $\ell$ as the muon ($\mu$), $E_{\rm min} \simeq \tfrac1{10} E_\mu$, and $s\simeq 2 E_\nu m_N$ ($m_N$ is the nucleon mass) with $E_\nu=10~{\rm TeV}$, we find $F_S\sim 0.9$. This implies that roughly $10\%$ of all events will contain some prompt real and energetic photon radiation. Since the photon is absorbed into the cascade rather than being associated with the track, this distorts the energy distributions of the charged lepton and the rest of the cascade. This can be compared to the probability of a high-energy muon undergoing hard bremsstrahlung within one radiation length, which is given parametrically by $(m_e/m_\mu)^2 \sim 2.4\times 10^{-5}$ \cite{ParticleDataGroup:2020ssz}.

In what follows we will discuss QED radiative corrections in their simplest form ({\it cf.} the discussion in Refs.~\cite{DeRujula:1979grv, Sigl:1997cq}), focusing on FSR. We make use of splitting functions, working at fixed order, and capturing the doubly-logarithmically enhanced corrections. These are the most important radiative corrections, and are large enough to impact current HE and UHE neutrino experiments. A full $O(\alpha)$ account of radiative corrections demands a one-loop electroweak calculation, in addition to the treatment of real final state photons \cite{Marciano:1980pb, Sirlin:1981yz, Sarantakos:1982bp, Wheater:1982yk, Bahcall:1995mm, Arbuzov:2004zr, Diener:2003ss, Tomalak:2021hec, Tomalak:2022xup}. This level of accuracy ($\sim 1\%$) is, however, extraneous for neutrino telescopes given their current statistical samples, and detector resolutions. Generalizations beyond our fixed-order treatment, i.e., to include leading-log resummation are straightforward, but unnecessary for applications where a $\sim 10\%$ accuracy is sufficient. Since our focus is on QED rather than QCD, we will always assume that the QCD-based neutrino CCDIS cross section is given. We make use of toy parametrizations for illustration, and realistic QCD calculations on the isoscalar target (i.e., averaging over proton and neutron)~\cite{Xie:2023suk} when discussing the numerical impact on observables.

The rest of the paper is organized as follows: in Sec.~\ref{sec:splitting}, we review relevant facts about neutrino interactions and QED splitting functions. 
In Sec.~\ref{sec:HE_nu}, we discuss the impact of FSR on the energies of the neutrino-interaction final states. We also discuss the impacts including inelasticity distributions, distinguishing neutrinos and antineutrinos, throughgoing muons, and double-bang signatures in HE neutrino telescopes. 
In Sec.~\ref{sec:UHE_nu}, we discuss UHE neutrino detection and observables that are impacted by FSR, including flavor measurements, Earth-emergent tau lepton, and $\nu_\tau$ regeneration. 
In Sec.~\ref{sec_SED}, we discuss FSR’s impacts on neutrino flux and spectrum measurements.
In Sec.~\ref{sec:FPF}, we comment on similar phenomenology at the FPF. 
Finally, in Sec.~\ref{sec:conclusions}, we summarize our findings.

\section{Charged-current neutrino scattering and final state radiation \label{sec:splitting}}

In what follows, we will have in mind charged-current (CC) neutrino (or antineutrino) scattering on nucleons. We therefore consider the interaction 
\begin{equation}
    {\nu}_\ell + N \rightarrow  X  + (\ell^- + n\gamma)~,
\end{equation}
where $N$ is a nucleon, $X$ is any hadronic final state, and we have allowed for $n$ final state photons emitted from the charged lepton. (Note that the photon emission from quarks is not important because the detectors cannot distinguish the hadronic shower and electromagnetic shower. Photon emission from the intermediate $W$ boson is also negligible because it is suppressed by the heavy $W$ mass.)
In this paper, we consider $n=0$ and $n=1$. Convenient kinematic variables are Bjorken $x$ and the inelasticity $y$, 
\begin{align}
    x&= \frac{Q^2}{2p_N\cdot Q} = \frac{(p_X - p_N)^2}{2p_N\cdot(p_X-p_N)}~, \\
    y&= \frac{E_X}{E_\nu}  = \frac{p_N\cdot E_X}{p_N \cdot p_\nu} ~. 
\end{align}
Notice that both variables can be defined independent of the charged lepton or photon kinematics. For tree-level scattering, suppressing terms of order $M/E_\nu$ and $m_\ell/E_\nu$,  the cross section can be written as \cite{Formaggio:2012cpf}
\begin{equation}
    \begin{split}
    \label{dsigma-dxy}
    \frac{\dd^2 \sigma^{(0)}_{\nu,\overline{\nu}}}{\dd x \dd y} =&\frac{ G_F M E_\nu }{\pi (1+ Q^2/M_W^2)^2} \\
    &\times \bigg[ y^2 F_1 + (1-y)F_2\pm xy(1-y/2) F_3\bigg]~,
    \end{split}
\end{equation}
where $F_i = F_i(x,Q^2)$ are structure functions and the ``$+$'' is for $\nu$ and the ``$-$'' is for $\bar{\nu}$. Predictions for the DIS cross sections can be found in, e.g., Refs.~\cite{Gandhi:1995tf, Gandhi:1998ri, Cooper-Sarkar:2011jtt, Connolly:2011vc, Chen:2013dza, Bertone:2018dse, Xie:2023suk}. In what follows, we only need $\dd \sigma^{(0)}/\dd y$. 

Let us next consider $n=1$ photon in the final state. At leading-logarithmic accuracy, FSR factorizes leg-by-leg. Any radiation from the hadronic parts of the diagram will be captured in the cascade. Since the cascade is inclusive with respect to hadronic + electromagnetic energy deposition, the Kinoshita-Lee-Nauenberg (KLN) theorem \cite{Kinoshita:1962ur, Lee:1964is} guarantees that hadronic FSR does not generate any logarithmically enhanced QED radiative corrections. The only ``large'' QED effects are, therefore, those involving FSR off the charged-lepton leg, and these can be computed at leading-log accuracy using splitting functions.

Whenever the probability of emitting radiation is substantially less than one, we may obtain accurate estimates without a full resummation of the leading logarithms. For this purpose, a fixed-order calculation is sufficient. The relevant distribution function or $\ell \rightarrow \ell\gamma$ is given by \cite{Gribov:1972ri,Peskin:1995ev} 
\begin{equation}\label{splitting-fn}
P_{\ell \rightarrow \ell \gamma}(z) 
= 
\frac{\alpha}{2\pi } \log \left(\frac{s}{m_{\ell }^2}\right)  \left[ \frac{(1+z^2)}{\qty[1-z]_+}+\frac32 \delta(1-z)~ \right],
\end{equation}
where $(1-z)$ is the fraction of the charged lepton's momentum carried away by the photon. The ``$1/(1-z)$'' behavior\footnote{We use $1/(1-z)$ and $1/[1-z]_+$ interchangeably for this specific colloquial remark, but implement the plus distribution in our calculations.} 
of the splitting function is characteristic of bremsstrahlung and encodes the soft photon singularity.

The logarithmic enhancement in \cref{splitting-fn} stems from integrating over $ \int \dd p_T^2/p_T^2$ from $p_T \sim m_\ell$ to $p_T \sim s$. A detailed discussion of an ``improved'' approximation scheme can be found in Ref.~\cite{DeRujula:1979grv} (replacing $s$ by $s(1-y+xy)$ with $x$ the Feynman $x$ and $y$ the inelasticity). However, at leading-log accuracy, this improvement does not affect the results. For simplicity, we use \cref{splitting-fn} as written, but refer the interested reader to Section 5 of Ref.~\cite{DeRujula:1979grv} for more details. 
Note that at leading-log accuracy $\log(\tfrac{s(1-y+xy)}{m^2}) \simeq \log(\tfrac{s}{m^2})$ for high and ultrahigh energy neutrinos.

\begin{figure}[t!]
    \centering
    \includegraphics[width=\linewidth]{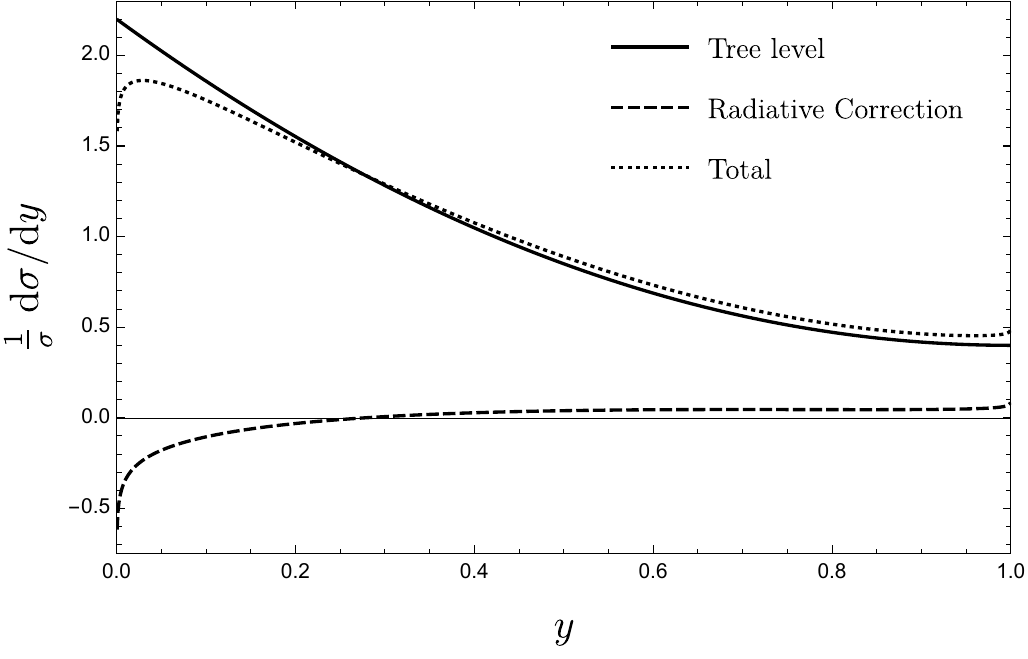}
    \caption{Illustration of the effect of FSR on the inelasticity distribution using \cref{toy-dist} for input parameters of $\langle y \rangle_{0}=0.35$,  and $\lambda_{0}=1.00$, and $E_\nu=100~{\rm TeV}$.  In the leading-log approximation, the effect of FSR is to migrate strength from smaller $y$ to larger $y$ without influencing the normalization of the distribution. \label{fig_realistic_cont}}
\end{figure}

Let us write $\dd \sigma = \dd \sigma^{(0)} + \dd \sigma^{(1)} + \ldots$, where $\dd \sigma^{(1)}$ includes all $O(\alpha)$ corrections to the differential cross section. The correction to the cross section contains a piece due to internal bremsstrahlung, and a virtual correction. These pieces conspire to ensure that inclusive observables (e.g., at fixed hadronic energy transfer) contain no large kinematic logarithms.  If we consider a leptonic variable, for example, $E_\ell$, then the $O(\alpha)$ logarithmically enhanced corrections to the cross section are given by 
\begin{equation}
    \label{sigma-1}
    \begin{split}
    \frac{\dd \sigma^{(1)}}{\dd E_\ell} = \frac{\alpha}{2\pi}& \int \dd y\int \dd z ~ \frac{\dd \sigma^{(0)}}{\dd y}\delta(E_\ell - (1-y) z E_\nu)\\
    &~\times  \log \left(\frac{s}{m_{\ell }^2}\right) \qty[\frac{1+z^2}{[1-z]_+} + \frac32 \delta(1-z)]~. 
    \end{split}
\end{equation}
This quantity is infrared (IR)-safe (but {\it not collinear} safe). When considering distributions binned as a function of $E_\ell$,  then the bin width in the charged-lepton energy serves as an effective IR-cutoff scale in estimating the size of double-logarithmic enhancements. In our numerical estimates we take $\alpha = \alpha(M_Z)$ with $\alpha^{-1}(M_Z)=129$ \cite{CERN:1995gra}.

Before proceeding, let us note that more sophisticated techniques have been developed for the resummation of final-state photons. Standard parton shower tools can automatically include final state radiation provided that the appropriate settings are ``turned on''. In what follows, we will use \cref{sigma-1} for all of our numerical illustrations, and fixed-order treatments are likely sufficient for most (if not all) applications at neutrino telescopes. If $\sim 1\%$ accuracy is required, then a full QED calculation is warranted (see, e.g., \cite{Tomalak:2021hec,Tomalak:2022xup} for a discussion in the context of GeV neutrino scattering); many $O(\alpha)$ effects (without logarithmic enhancement) are not captured by parton showers. Nevertheless, if it proves more convenient in experimental analysis, \cref{sigma-1} can be replaced with an equivalent parton shower description.

For illustration's sake, let us make use of IceCube's simplified parametrization 
of $\dd \sigma^{(0)}/\dd y$~\cite{IceCube:2018pgc}, 
\begin{equation}
    \begin{split}
    \label{toy-dist}
    \frac{1}{\sigma^{(0)}}\frac{\dd \sigma^{(0)}}{\dd y}
    &= ~C(\epsilon_0,\lambda_0) (1+\epsilon(1-y)^2) y^{\lambda_0-1} ~,
    \end{split}
\end{equation}
where $\epsilon_0$ is in practice determined by specifying the mean inelasticity $\langle y \rangle_0$ and using the formula 
\begin{equation}
    \epsilon_0 =-\frac{((\lambda_0 +2) (\lambda_0 +3)) ((\lambda +1) \langle y \rangle_0 -\lambda_0)}{2 ((\lambda_0 +3) \langle y \rangle_0-\lambda_0 )}~. 
\end{equation}
The constant $C$ is fixed by normalization.

In \cref{fig_realistic_cont} we show the effect of FSR computed using \cref{sigma-1} for input parameters of $\langle y \rangle_{0}=0.35$ and $\lambda_{0}=1.00$; these input parameters correspond to realistic choices for $E_\nu \simeq 100~{\rm TeV}$~\cite{IceCube:2018pgc}. 
Having discussed the general formalism we use (i.e., splitting functions at fixed order in $\alpha$) and identified corrections that are sizeable, we now turn to applications at neutrino telescopes. 

\section{High-energy neutrino observation \label{sec:HE_nu} }
In this section, we discuss how FSR impacts the observation of HE neutrinos ($100~{\rm GeV} \lesssim E_\nu \lesssim 100~{\rm PeV}$). 
High-energy neutrino telescopes detect neutrinos via two basic topologies: ``tracks'' and ``cascades'' ({\it cf.} \cref{cartoon}).
Tracks are formed by muons, which travel macroscopic distances.
Cascades include hadronic and electromagnetic energy deposition in a shower that is localized over a few hadronic interaction lengths. 
Events produced by HE neutrinos can originate from either inside or outside (i.e., throughgoing muons) the detector. 
When FSR is emitted, it transfers energy from the track part of the topology to the cascade. More generally, for instance, in the case of the ``double-bang'' signature that is used to detect tau neutrinos, FSR distorts the ratio of leptonic and hadronic energy estimators.

Therefore, in what follows, we distinguish two parts of energy: the \textit{charged-lepton energy}, $E_\ell$, and the \textit{shower energy}, $E_{\rm shower}$, which combines the energies of the FSR photon and hadronic shower.\!\footnote{The hadronic shower also contains a large amount of electromagnetic activity due to, e.g., $\pi^0\rightarrow \gamma \gamma$).} The sum, $E_{\rm shower}+ E_\ell$, can be used as an estimator for the neutrino energy $E_\nu$. Whenever an observable measures $E_{\rm shower}$ or $E_\ell$ separately, however, QED corrections can be enhanced by large FSR kinematic logarithms and be sizeable, as shown below. 

In \cref{Fig_deltaE_Enu}, we show the average relative shift in the charged-lepton energy ($\langle \Delta E_\ell \rangle/E_\ell$) and the shower energy ($\langle \Delta E_{\rm shower} \rangle/E_{\rm shower}$) due to FSR. By definition, $\langle \Delta E_\ell \rangle + \langle \Delta E_{\rm shower} \rangle = 0$. The calculation is based on realistic neutrino CCDIS $\dd \sigma^{(0)}/\dd y$ distributions based on QCD and the isoscalar target~\cite{Xie:2023suk, Keping_prviate} (the DIS differential cross sections we use as input and after FSR correction can be found \href{https://github.com/beizhouphys/DIS_dsigma-dy_with_and_without_FSR}{at Github} \github{beizhouphys/DIS_dsigma-dy_with_and_without_FSR}). 
For leptons, the shifts can be as large as $\sim 5\%$ whereas for the shower energy the estimated energy can shift by as much as $\sim 25\%$(!); this is a consequence of the tree-level distribution being asymmetric between the leptonic and hadronic energy.
Moreover, it is important to note that the shifts in the shower energies will be further enhanced by $\simeq 10-20\%$ in the realistic experimental settings, because electromagnetic showers have more light yields than hadronic showers~\cite{IceCube:2013dkx}.
Therefore, Fig.~\ref{Fig_deltaE_Enu} demonstrates that QED FSR can substantially distort experimental observables at a level that is relevant for existing neutrino telescopes. 
Notice that the relative energy shifts increase with energy, which reflects the increased probability of FSR for high- vs. low-energy leptons.  Convenient scaling relations for charged-lepton energies are given as follows:
\begin{align}
    \frac{\langle \Delta E_\mu \rangle}{E_\mu}
    &\simeq 
    4.6\% + 0.0075 \times \log_{10} \left(\frac{E_\nu}{10^{10}\, {\rm GeV}}\right)  \label{eq_deltaEmu},\\
    \frac{\langle \Delta E_\tau \rangle}{E_\tau} &\simeq 
    3.7\% + 0.0075 \times \log_{10} \left(\frac{E_\nu}{10^{10}\, {\rm GeV}}\right) ~.
     \label{eq_deltaEtau}
\end{align}
The effects of FSR are enhanced for muons relative to tau leptons due to the $\log(s/m_\ell^2)$ collinear enhancement. Finally, notice that there is a difference between $\nu$ and $\bar{\nu}$, which we will comment on in \cref{sec_HEnu_inel_ratio}. 

Finally, one can get the scaling relations for shower energies through $E_\ell + E_{\rm shower} = E_\nu$ and $\Delta E_\ell + \Delta E_{\rm shower} = 0$:
\begin{equation}
\frac{ \langle \Delta E_{\rm sh} \rangle}{E_{\rm sh}} = - \frac{1-y}{y} \frac{\langle \Delta E_\ell \rangle}{E_\ell}, 
\end{equation}
where $\ell = \mu$ or $\tau$ and the corresponding $\frac{\langle \Delta E_\ell \rangle}{E_\ell}$ are given in Eqs.~\eqref{eq_deltaEmu} and \eqref{eq_deltaEtau}.

\subsection{Inelasticity measurements 
and neutrino telescopes
\label{sec_HEnu_inel} }
As we have discussed above, a simple exclusive observable that is sensitive to FSR is the differential cross section $\dd \sigma/\dd E_\mu$ in muon-neutrino CC scattering. At leading order in $\alpha$ (i.e., without FSR) this observable is trivially related to the inelasticity distribution $\dd \sigma^{(0)}/ \dd y$. As is immediately evident from \cref{sigma-1}, this simple correspondence is violated once FSR is included. The experimental definition used as an estimate of the inelasticity is 
\begin{equation}
    y_{\rm exp} 
    \equiv
    \frac{E_{\rm shower}}{E_{\rm track}+E_{\rm shower}}
    = y_{\rm QCD} + \frac{E_\gamma}{E_\nu}\,,
\label{eq_yexp}
\end{equation}
where $y_{\rm QCD}=E_{X}/E_\nu$ and $E_\gamma$ is the energy deposited by FSR from the lepton leg. 

\begin{figure}[t]
\includegraphics[width=0.49\textwidth]{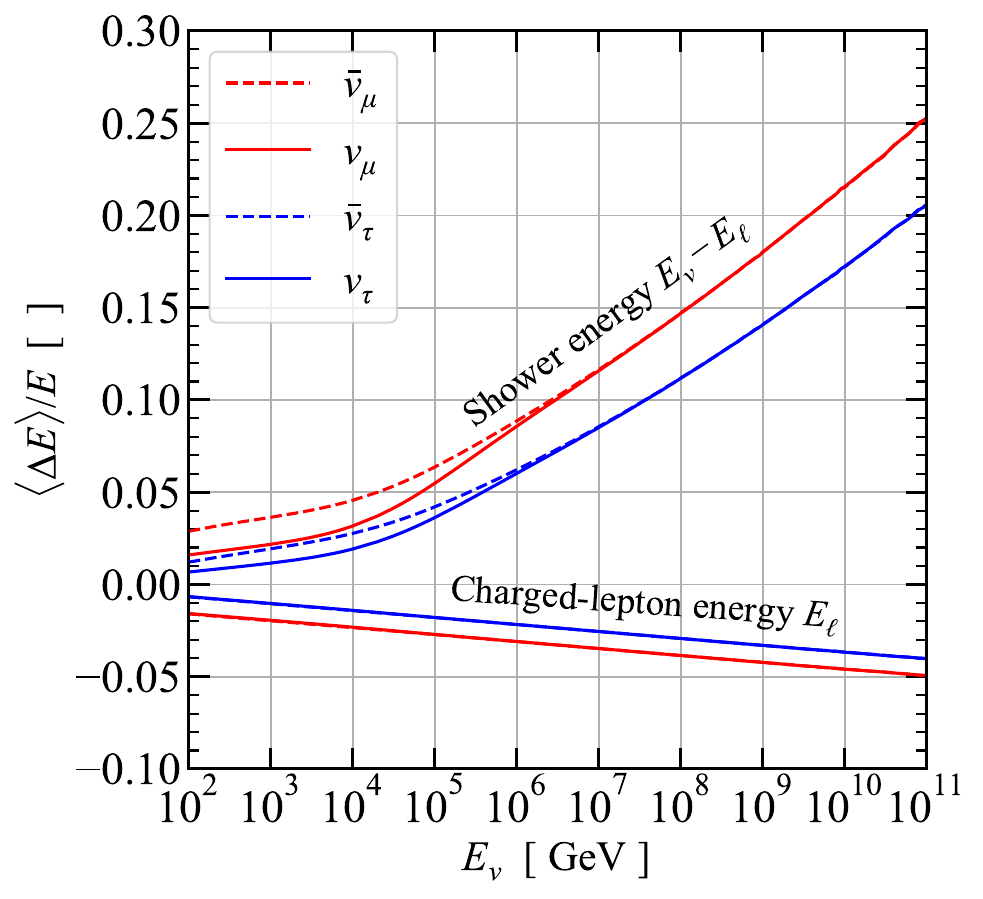}
\caption{Average relative shift in energy due to FSR for the final-state charged-lepton energy (\textbf{curves below zero}) and shower energy, i.e., ``the rest'' $E_\nu-E_\ell$  (\textbf{curves above zero}) from neutrino CCDIS. For example, taking a HE muon neutrino in IceCube the lepton energy corresponds to the track, and the rest corresponds to the shower. Curves are plotted as a function of the parent neutrino energy ($E_\nu$).
}
\label{Fig_deltaE_Enu}
\end{figure}

\begin{figure*}[t!!]
\includegraphics[width=0.49\textwidth]{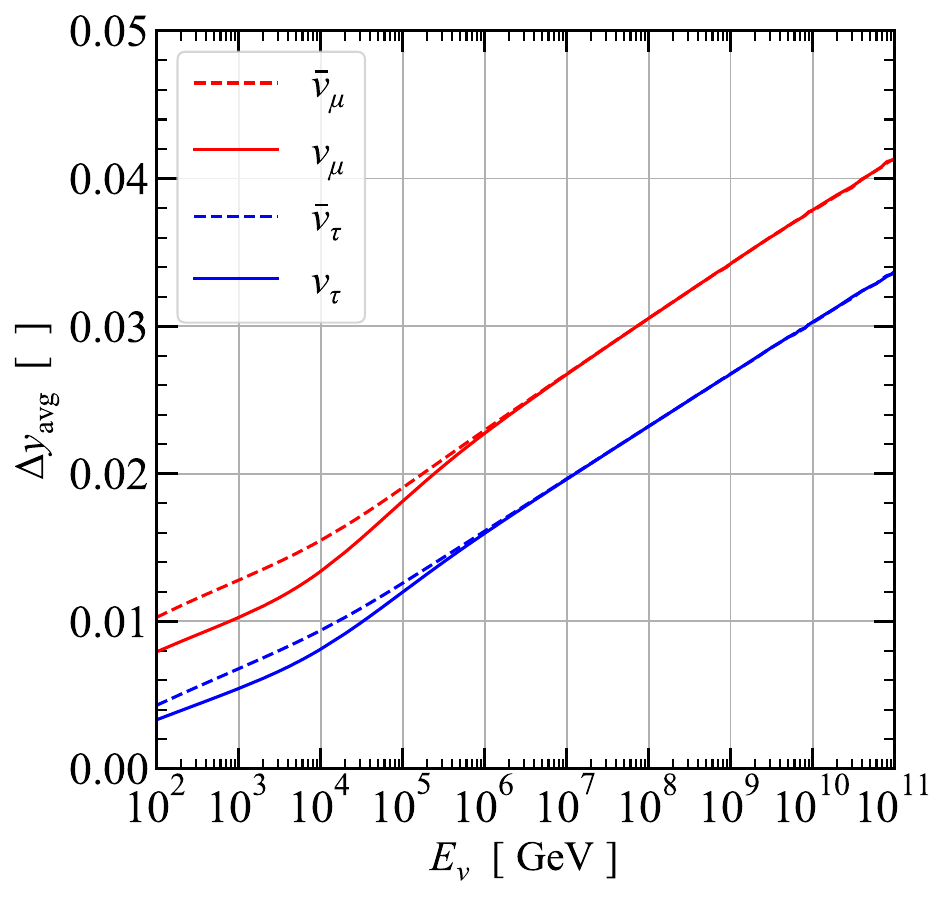}
\includegraphics[width=0.472\textwidth]{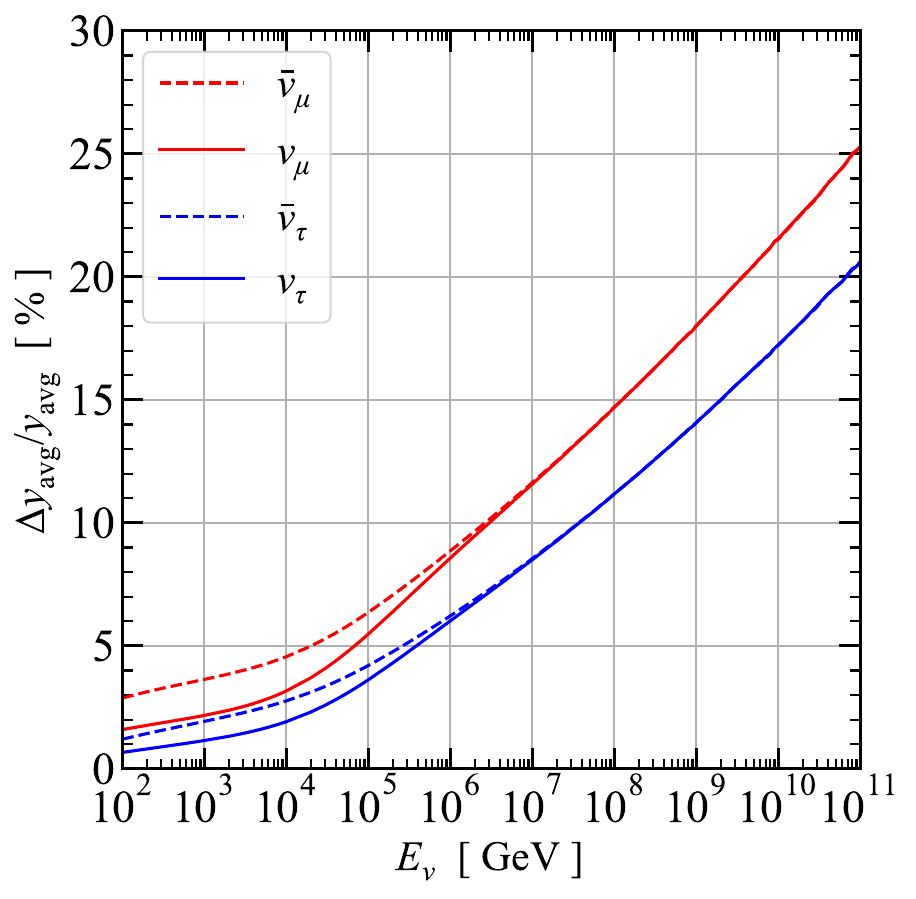}
\caption{Shifts in the experimentally measured average inelasticity, $y_{\rm avg} = 1 - \langle E_\mu\rangle / E_\nu$, from neutrino CCDIS due to the FSR photon taking part of the energy from the charged lepton to the hadronic side, as a function of the parent neutrino energy ($E_\nu$).
\textbf{Left panel:} the shift, $\Delta y_{\rm avg} \equiv \langle E_\gamma \rangle / E_\nu$. \textbf{Right panel:} the relative shift, $\Delta y_{\rm avg} / y_{\rm avg}$. We do not include experimental efficiencies as a function of $y$ in the estimate of $y_{\rm avg}$~\cite{IceCube:2018pgc}. We also do not account for the difference in light yield between hadronic and electromagnetic showers, which will further enhance the shifts by $\simeq 10$--$20\%$~\cite{IceCube:2013dkx}. 
}
\label{Fig_DeltaYavg_Enu}
\end{figure*}

In \cref{Fig_DeltaYavg_Enu} we plot the absolute ($\Delta y_{\rm avg}$; left panel) and relative ($\Delta y_{\rm avg}/y_{\rm avg}$; right panel) shifts in the average experimental inelasticity $\langle y_{\rm expt} \rangle$, defined as $\Delta y_{\rm avg} \equiv \langle y_{\rm exp} \rangle - \langle y_{\rm QCD} \rangle = \langle E_\gamma \rangle/E_\nu$. This is a useful quantitative measure of FSR. 
The trends of the curves are related to that of \cref{Fig_deltaE_Enu}. 
Since $\langle E_\gamma \rangle >0$, $\Delta y_{\rm avg}$ is always positive.  
The increase in $\Delta y_{\rm avg}$ with $E_\nu$ occurs both because of increasing logarithmic enhancements $\log(s/m_\ell^2)$, and because $\langle y_{\rm QCD}\rangle$ decreases as $E_\nu$ increases. The relative shift due to FSR, $\Delta y_{\rm avg}/y_{\rm avg}$, can be as large as 25\%(!).
Again, it is important to note that the shifts in $y_{\rm avg}$ will be further enhanced by $\simeq 10-20\%$ in the realistic experimental settings, because electromagnetic showers have more light yields than hadronic showers~\cite{IceCube:2013dkx}.
The difference in the light yield also affects the inference of the parent neutrino energy from the measured total energy, although it is minor.

Both \cref{Fig_deltaE_Enu,Fig_DeltaYavg_Enu} demonstrate that the impact of FSR is substantial, and influences observables at a level that is relevant for ongoing and near-term experiments. This is important because the inelasticity distribution has many useful applications. For instance, the differing dependence of \cref{dsigma-dxy} on $y$ for $\nu$ vs. $\bar{\nu}$ allows one to statistically infer the flux ratio $\Phi_\nu / \Phi_{\bar{\nu}}$ which has interesting applications both for astrophysics and for inferring neutrino mixing parameters. In addition to application to astrophysics and neutrino physics, the inelasticity distribution also provides interesting constraints on QCD distributions such as the charm quark fraction of nucleons. In what follows (\cref{sec_HEnu_inel_ratio,sec_HEnu_inel_fundamental}) we discuss the impact of FSR on these applications in turn.

\subsubsection{Measuring \texorpdfstring{$\nu/\bar{\nu}$}{nu/nubar} flux ratio}
\label{sec_HEnu_inel_ratio}
The ratio of neutrinos to antineutrinos is strongly correlated with the atmospheric muon charge ratio. The excess of $\mu^+$ vs. $\mu^-$ stems from the excess of protons in cosmic rays and the steeply falling cosmic ray spectrum \cite{Frazer:1972ep, Gaisser:2011klf}. Since neutrinos and muons both come from weak decays of hadrons (primarily $\pi^\pm$ and $K^\pm$), measurements of the $\nu/\bar{\nu}$ flux ratio provide complementary information on cosmic rays and an interesting cross-check on the $\mu^+$ to $\mu^-$ ratio. Furthermore, measuring the astrophysical $\nu/\bar{\nu}$ flux ratio is an important and powerful tool for testing the source properties and neutrino properties.

The inelasticity distribution is a useful discriminator of $\nu$ vs. $\bar{\nu}$ due to the sign difference multiplying the structure function $F_3$ in \cref{dsigma-dxy}.  
At the quark level, this is because neutrinos and antineutrinos are sensitive to different quark flavors with different densities, especially at large Bjorken $x$ values where valence quarks dominate, leading to different cross sections (e.g., Fig.~1 of Ref.~\cite{Zhou:2019vxt}) and inelasticity distributions (e.g., Fig.~8 of Ref.~\cite{IceCube:2018pgc}). A consequence of these differing inelasticity distributions is that FSR induces different shifts in $y_{\rm avg}$. 

We illustrate this point in \cref{Fig_DeltaYavg_Enu}, where the shift in $y_{\rm avg}$ is due to FSR. 
The difference between $\nu$ and $\bar{\nu}$ decreases as neutrino energy increases as smaller Bjorken $x$ is favored, where sea quarks are more dominant, and almost vanishes for $E_\nu \gtrsim 10^6$ GeV. Since atmospheric neutrino fluxes are larger than astrophysical ones,  only the atmospheric $\nu/\bar{\nu}$ flux ratio has been measured~\cite{IceCube:2018pgc}, but the astrophysical $\nu/\bar{\nu}$ flux ratio could potentially be measured in the future.

\begin{figure}[t!!]
\includegraphics[width=0.49\textwidth]{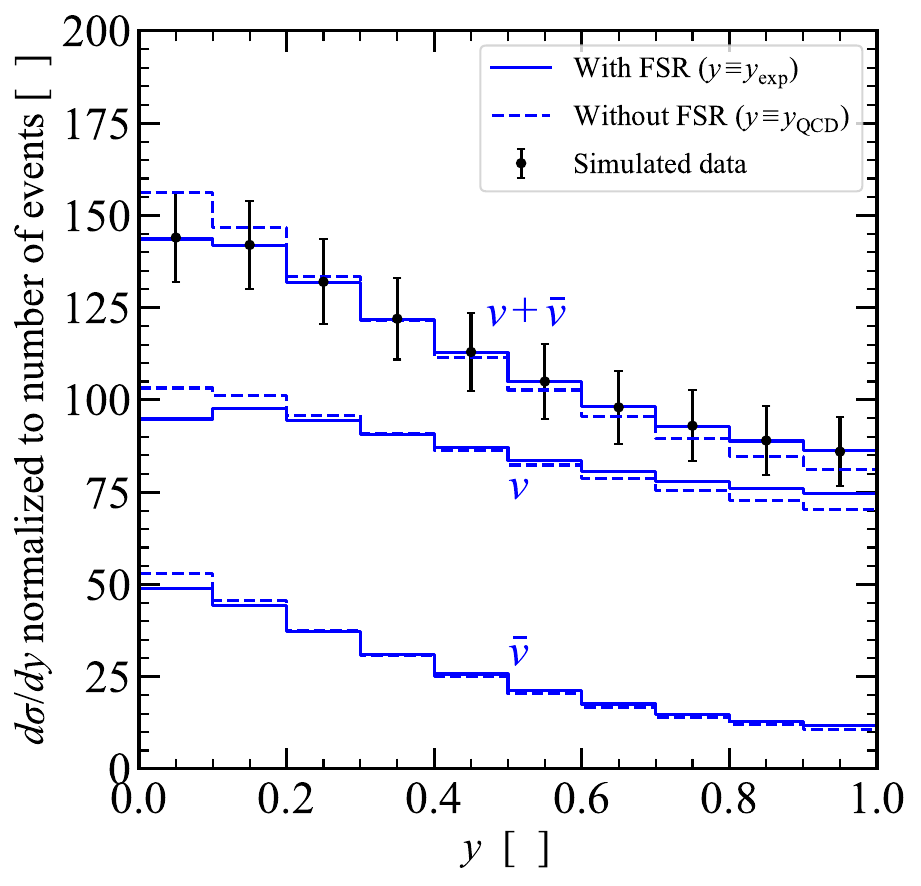}
\caption{
Illustration of how FSR can impact determinations of the $\nu/\bar{\nu}$ flux ratio. We have generated pseudo-data, based on the top solid histogram, that mimic the IceCube Collaboration's measurements of the $\nu/\bar{\nu}$ flux ratio using inelasticity distributions of starting muon events~\cite{IceCube:2018pgc}. 
The histograms are produced assuming the same statistics as in Ref.~\cite{IceCube:2018pgc}.
The dashed histograms are derived from QCD-based neutrino-isoscalar CCDIS inelasticity distributions~\cite{Xie:2023suk, Keping_prviate}, which are corrected by FSR (\cref{sigma-1}) to get the solid histograms.
}
\label{Fig_dsgmdy_data}
\end{figure}

In \cref{Fig_dsgmdy_data} we illustrate how IceCube collaboration measured the $\nu/\bar{\nu}$ flux ratio using the experimental inelasticity defined above~\cite{IceCube:2018pgc} and how FSR would affect the measurement. For each starting muon event, its inelasticity and total energy ($E_{\rm vis}$; close to the parent neutrino energy) are measured. 
The 2650 events collected over 5 years are distributed across five different $E_{\rm vis}$ bins (\cref{Fig_dsgmdy_data} only shows the lowest-energy bin), and within each $E_{\rm vis}$ bin, the events are further distributed into ten $y_{\rm exp}$ bins. Then, the $\nu/\bar{\nu}$ flux ratio is measured by fitting the data with theoretical predictions of $\nu$ and $\bar{\nu}$ components, including cross sections, $\dd \sigma / \dd y_{\rm QCD}$, and fluxes.
The histograms in \cref{Fig_dsgmdy_data} are generated purely from theory including FSR, with no detector effects included, and the data are simulated based on the same statistics of Ref.~\cite{IceCube:2018pgc}. 
In reality, statistics in the bins with the smallest and largest $y_{\rm exp}$ are suppressed due to detector effects (see Fig.~7 of Ref.~\cite{IceCube:2018pgc}). 
However, our Fig.~\ref{Fig_dsgmdy_data} shows that the largest distortion of FSR on extracting the $\nu/\bar{\nu}$ flux ratio also comes from the bins with the smallest and largest $y_{\rm exp}$ (see also \cref{fig_realistic_cont}). 
One could, in principle, perform a Poissonian likelihood fit to the data with and without FSR, and we find it gives unrealistic estimates.
Therefore, it is crucial to include the proprietary experimental information such as efficiencies, energy smearing, and missing energy to get a realistic estimate of FSR's impact on extracting the $\nu/\bar{\nu}$ flux ratio.

Here, we attempt a simplified fit (but less sensitive to the experimental details) of the $\nu/\bar{\nu}$ flux ratio using the mean inelasticity measurement plotted in Fig.~8 of Ref.~\cite{IceCube:2018pgc}. 
For the theoretical input, we take the CSMS $y_{\rm avg}(E_\nu)$ curves from the same figure, and multiply them by the factors given in \cref{Fig_DeltaYavg_Enu} to get the FSR corrected curves.
Next, we sum the $\nu$ and $\bar{\nu}$ curves weighted by the corresponding cross sections~\cite{Cooper-Sarkar:2011jtt} and atmospheric neutrino fluxes,\!\footnote{The atmospheric neutrino fluxes are calculated using the Hillas-Gaisser H3a cosmic ray model~\cite{Gaisser:2013bla}, and the {\tt Sybill2.3c} hadronic interaction model~\cite{Riehn:2017mfm}.} 
with the $\nu/\bar{\nu}$ flux ratio as a free-floating parameter.
We then extract the $\nu/\bar{\nu}$ flux ratio by fitting the theory, for the cases without FSR and with FSR, to the data. The extracted value of the $\nu/\bar{\nu}$ flux ratio is then divided by the input value of the $\bar{\nu}/\nu$ ratio (``prediction''), i.e., $R_{\rm fit} = (\nu/\bar{\nu})_{\rm fit} \big/ (\nu/\bar{\nu})_{\rm prediction}$.
We find that FSR shifts the inferred $\nu/\bar{\nu}$ flux ratio, $R_{\rm fit}$, by about $5\%$ relative to the case of not including FSR. This is to be compared with $0.77^{+0.44}_{-0.25}$ from the previous IceCube measurement~\cite{IceCube:2018pgc}. 
We note again that it is necessary to include the experimental effects to get a more rigorous estimate and direct comparison with the previous measurement.
The experimental uncertainty will decrease quickly thanks to the increasing data and especially future telescopes (e.g.,  IceCube-Gen2~\cite{IceCube-Gen2:2020qha}) that probe higher energies where FSR's effect is more prominent.

\subsubsection{Applications to fundamental physics}
\label{sec_HEnu_inel_fundamental}
The $\nu/\bar{\nu}$ flux ratio is also interesting from the perspective of both neutrino oscillation and hadronic physics. Inelasticity is a useful tool for constraining the mass hierarchy, the atmospheric mixing angle $\theta_{23}$, and the {\it CP}-violating phase $\delta_{\rm CP}$ \cite{Ribordy:2013xea, Ge:2013ffa}.  It also gives valuable information on the strange quark parton-distribution functions (PDFs) of nucleons \cite{IceCube:2018pgc}. 

Reference~\cite{Ribordy:2013xea} concludes that a complete $\nu/\bar{\nu}$ separation would enhance sensitivity to CP-violation and the mass hierarchy by a factor of 2--3, however detector resolution effects limit the improvement in practice to $\sim 1.25$ i.e., $25\%$. Ref.~\cite{Ge:2013ffa} reaches similar conclusions estimating the ultimate impact of inelasticity separation as a $\sim 40\%$ improvement in the sensitivity to the mass hierarchy. This issue has been recently revisited in Ref.~\cite{Olavarrieta:2024eaq}, where the authors reach similar conclusions. As we discussed above, the asymmetric nature of $\dd \sigma/\dd y$ means that the $\nu/\bar{\nu}$ flux ratio is unusually sensitive to radiative corrections and so FSR may affect extractions of neutrino oscillation parameters from atmospheric neutrino data.

Another application of inelasticity to fundamental physics is statistical charm tagging. Strange sea quarks give rise to sizeable charm production rates, since these are proportional to the Cabibbo-Kobayashi-Maskawa matrix elements $|V_{cs}|^2\gg |V_{uc}|^2$. Cutting on the inelasticity can supply ``charm-rich'' samples. For example, IceCube's analysis estimates that more than $1/3$ of all events with $y_{\rm expt} >0.8$ contain charm mesons~\cite{IceCube:2018pgc}. The effect of FSR is to shuffle energy from the leptonic system into the hadronic system, thereby increasing $y$ for all CCDIS events. Therefore, when attempting to statistically identify charm production rates it is important to account for increased valence quark contributions that ``leak'' into the signal window due to FSR. A related discussion of charm production in the context of more granular detectors at the LHC's FPF is presented further below. 

\subsection{Throughgoing muons and parent neutrino energy} 
\label{sec_HEnu_thrgo}

In the above discussion, we have largely focused on events in which shower and track energies are separately observable. This is a good description of so-called ``starting'' events, but most HE neutrino events originate outside the active volume of the detector. In this case, any FSR would be absorbed outside the detector, along with the hadronic shower, and be missing from the energy budget. The effect of QED FSR is therefore to systematically bias muon tracks to lower energies. 

Throughgoing muon tracks are associated with muons produced externally to the detector, but pass through the active volume leaving a track of Cherenkov radiation. Such events are important for high-energy neutrinos since the fluxes at these energies are much smaller, and the effective volume over which a throughgoing muon can be produced is an order of magnitude larger than the active volume of the detector. Including FSR will lower the energy of throughgoing muons by as much as $\sim 5\%$, and should be included when estimating the parent neutrino energy. 

\subsection{Impact on the double-bang signature from $\nu_\tau$}
In addition to the muon track events discussed above, another measurement impacted by FSR is the ``double-bang'' signature. This is induced by CCDIS of $\nu_\tau/\bar{\nu}_\tau$ at $E_\nu\gtrsim 10^5~{\rm TeV}$, with the first bang/shower formed by the hadronic cascade and the second (separated) bang/shower formed by the tau lepton decay~\cite{Learned:1994wg, IceCube:2020fpi}.

When FSR happens on $\nu_\tau/\bar{\nu}_\tau$ CCDIS, the photon takes energy from the tau lepton (second bang) to the hadronic cascade (first bang) (\cref{Fig_deltaE_Enu}). The distortion in the energy balance between the two bangs will distort the inference of the parent $\nu_\tau/\bar{\nu}_\tau$ energy. Moreover, a reduction of the tau lepton energy also decreases the separation between the two bangs, both spatially and temporally, which would make it harder to identify the double-bang signature.

\bigskip

Before discussing UHE detection in the next section, we briefly comment on the detection of $\nu_e/\bar{\nu}_e$ CC events.
Incident $\nu_e$ and $\bar{\nu}_e$ produce high-energy $e^\mp$. Unlike muons, electrons travel relatively short distances (on the order of the radiation length $X_0\sim 30~{\rm cm}$) and are contained in the same cascade as the hadronic shower. Since photons initiate electromagnetic showers over the same length scale, both the electron {\it and} the photon are contained within a single cascade. These events are therefore inclusive with respect to FSR and do not have any large kinematic logarithms as is guaranteed by the KLN theorem. For the same reasons, FSR from $\nu_e/\bar{\nu}_e$ CC does not affect the balance between hadronic and lepton showers~\cite{Li:2016kra}.

\section{Ultrahigh-energy neutrino observations \label{sec:UHE_nu} }

In this section, we discuss the impact of FSR on UHE neutrino observation. Most of the discussions are based on our results in \cref{Fig_deltaE_Enu,eq_deltaEmu,eq_deltaEtau}.

There are two major observational strategies for the detection of UHE neutrinos~\cite{Ackermann:2022rqc}. The first approach involves measurements of coherent Askaryan radiation (nanosecond radio flashes) emitted by the electromagnetic showers generated from UHE neutrino interactions in dense media, such as ice~\cite{Askaryan:1961pfb, Ackermann:2022rqc}. Experiments pursuing this detection strategy include ANITA~\cite{ANITA:2008mzi}, ARIANNA~\cite{ARIANNA:2019scz}, PUEO~\cite{PUEO:2020bnn}, the IceCube-Gen2 radio array~\cite{IceCube-Gen2:2020qha}, etc.
The second strategy is to detect Earth-emergent charged (mainly tau) leptons from UHE neutrino interactions within the Earth or other material outside the actively instrumented detector volume. Experiments pursuing this approach are PUEO~\cite{PUEO:2020bnn}, POEMMA~\cite{POEMMA:2020ykm}, TAMBO~\cite{Romero-Wolf:2020pzh}, GRAND~\cite{GRAND:2018iaj}, etc.
For a comprehensive overview of UHE neutrino experiments, see Ref.~\cite{Ackermann:2022rqc}.

\subsection{Askaryan radio detectors}
\label{sec_UHEnu_Askaryan} 

\subsubsection{Overall detectable energy}
\label{sec_UHEnu_Askaryan_DetEn} 

For the in-ice Askaryan radio detectors, FSR increases the overall detectable energy of the detector, which also effectively lowers the detection energy threshold. 
For $\nu_\tau$/$\bar{\nu}_\tau$ CCDIS, the energy loss of the tau lepton is not as strong as the muon or electron due to its heavy mass, and it could deposit only a small fraction or even none of its energy to the detector~\cite{Christoph_prviate}. Therefore, FSR could increase the overall detectable energy of $\nu_\tau$/$\bar{\nu}_\tau$ CCDIS by up to 20\% (Fig.~\ref{Fig_deltaE_Enu}), as the photon deposits nearly all of its energy in the detector.
For $\nu_\mu$/$\bar{\nu}_\mu$ CCDIS, the enhancement from FSR is less, as the muon energy loss is stronger than $\tau$, and a detector-level simulation is needed to quantify the effect. For $\nu_e$/$\bar{\nu}_e$ CCDIS, there is no enhancement from FSR, as the electrons also deposit nearly all the energy in the detector.

\subsubsection{Flavor measurement}
Recent work has explored the possibility of using a detailed substructure of Askaryan radiation to separate charged lepton and hadronic showers from one another~\cite{Testagrossa:2023ukh, Coleman:2024scd}. 
The new method proposed in Ref.~\cite{Coleman:2024scd} involves measuring the combined $\nu_\mu\&\bar{\nu}_\mu+\nu_\tau\&\bar{\nu}_\tau$ flux by identifying events with at least one displaced shower (generated by tau leptons or muons while propagating in ice) relative to the primary shower from the neutrino CC interaction vertex.
A complementary strategy to identify $\nu_e\&\bar{\nu}_e$-induced events was also explored, wherein one takes advantage of the elongated shower produced by an electron from the $\nu_e$ CC interaction due to the Landau-Pomeranchuk-Migdal effect. The observable signal would then involve several slightly displaced sub-showers.
Collectively, these features elongate the particle profile of the shower and its radio emission, compared to the more compact hadronic showers generated in other types of interactions (e.g., neutral current events).

Both the $\nu_\mu\&\bar{\nu}_\mu+\nu_\tau\&\bar{\nu}_\tau$ and $\nu_e\&\bar{\nu}_e$ detection channels are affected by FSR. 
For muon and tau neutrinos, FSR reduces the detectability of displaced showers by reducing the energy of the charged lepton (see Fig.~3 of Ref.~\cite{Coleman:2024scd}). As a concrete example, the detection efficiency curves in Fig.~3 of Ref.~\cite{Coleman:2024scd}  would shift to the right by $\simeq 3\%$--$5\%$, depending on the neutrino energy and flavor. 
For the $\nu_e\&\bar{\nu}_e$ channel, FSR does not impact the signal because both electrons and photons initiate electromagnetic showers. However, FSR increases the rate of the background events originating from the CC interactions of $\nu_\mu/\bar{\nu}_\mu/\nu_\tau/\bar{\nu}_\tau$ with the sub-shower (from muon/tau lepton) very close to the primary shower, thus mimicking the signal. 
Without considering FSR, Ref.~\cite{Coleman:2024scd} concludes that the background rate is negligible. With the inclusion of FSR, the photon would create a detectable sub-shower by itself, leading to a higher background rate. Moreover, the method proposed in Ref.~\cite{Coleman:2024scd} relies on ``...subtle features in the waveform shape...'' that are classified with a deep neural network. Elongated showers induced by FSR photons may distort these waveforms and introduce an uncontrolled systematic uncertainty. The relevant infrared cutoff is the energy at which a photon would induce a Landau–Pomeranchuk–Migdal (LPM) elongated shower. We take this threshold as $E_{\rm LPM}=3\times10^{16} ~{\rm eV}$ based on Refs.~\cite{Coleman:2024scd,ParticleDataGroup:2020ssz}. Using \cref{eq_sudakov} we find a  $\sim 15\%$ probability of FSR per flavor for a $3\times 10^{17} {\rm eV}$ neutrino; the effect grows proportional to $\log(E_\nu/E_{\rm LPM})$. 
Assuming comparable fluxes of $\nu_\tau:\nu_\mu:\nu_e$ then gives a $\sim 30\%$ background from $\nu_\mu/\bar{\nu}_\mu/\nu_\tau/\bar{\nu}_\tau$ interactions. 
Finally, it would also be useful to include
photonuclear interaction, which is larger than pair production above $10^{20}$~eV, and direct pair production and electronuclear interaction, which are larger than bremsstrahlung above $10^{20}$~eV~\cite{Gerhardt:2010bj}.

\subsection{Air shower detectors}
\label{sec_UHEnu_AirShower}
For the detection strategy involving Earth-emergent leptons, the impacts of FSR are straightforward to estimate. 
As the FSR photon and the hadronic shower are both absorbed in the Earth or mountain where the neutrino interacts, the detected charged-lepton energy will decrease by the amount given in \cref{eq_deltaEmu,eq_deltaEtau}. This process is analogous to the throughgoing muon track in HE neutrino telescopes.
Not including the FSR then leads to an underestimation of the parent neutrino energy by $\sim 5\%$. Although this is smaller than the energy resolution of upcoming UHE neutrino detectors (e.g., Refs.~\cite{GRAND:2018iaj, Glaser:2019cws, PUEO:2020bnn, Romero-Wolf:2020pzh, POEMMA:2020ykm}), this is a guaranteed correction from the Standard Model so it should be included in energy estimators sooner or later.

Finally, FSR also affects the $\nu_\tau$ regeneration process~\cite{Halzen:1998be, Bigas:2008sw, Alvarez-Muniz:2018owm}, in a manner similar to the Earth-emergent leptons discussed above. A new feature for $\nu_\tau$ regeneration is the possibility of multiple neutrino interactions, each of which may include FSR.\!\footnote{We note that FSR's impact on tau decay, $\tau \rightarrow \nu_\tau X + \gamma$, is not enhanced by a large logarithm, $\log(E_\nu/m_\tau)$,  since the Lorentz invariant energy scale is set by $m_\tau$.} Taking FSR into account {\it decreases} the regenerated $\nu_\tau$ flux by lowering the tau energy, which simultaneously {\it increases} the flux since lower-energy taus have a longer absorption length within the Earth. It would be interesting to incorporate FSR into simulations of $\nu_\tau$ regeneration and understand the interplay of these two effects.

\section{Implication on flux and spectrum measurements}
\label{sec_SED}

Any bias in the total detectable energy due to FSR discussed above will be amplified when measuring the neutrino flux due to the steeply falling spectrum, i.e., 
\begin{equation}
(1-\delta{E_\nu})^\Gamma \simeq 1 - \Gamma \times \delta{E_\nu},
\end{equation}
where $-\Gamma$ is the spectrum index and is usually $\leq -2$. That is to say, {\it not including FSR will underestimate the neutrino flux by $\Gamma \times \delta{E_\nu}$}.
For example, in the case of UHE $\nu_\tau$/$\bar{\nu}_\tau$ discussed in Sec.~\ref{sec_UHEnu_Askaryan_DetEn} and assuming $\Gamma=3$, the underestimation could be as much as $20\%\times3=60\%$. For another example, in the case of throughgoing muons~\cite{IceCube:2021uhz} (Sec.~\ref{sec_UHEnu_AirShower}) or Earth-emergent tau leptons (Sec.~\ref{sec_HEnu_thrgo}),  the underestimation could be as much as $5\% \times 3 \simeq 15\%$.

FSR also impacts the spectral shape measurements because its effects depend on the parent neutrino energy (Fig.~\ref{Fig_deltaE_Enu}). We estimate this impact as follows.
To measure a neutrino spectrum $F_\nu \propto E_\nu^{-\Gamma}$ between $E_{\nu,1}$ and $E_{\nu,2}$ ($E_{\nu,2} > E_{\nu,2}$), 
the spectral index, $-\Gamma$ could be determined by \begin{equation}
-\Gamma = \frac{\log_{10} F_{\nu, 2} - \log_{10} F_{\nu, 1}}{\log_{10} E_{\nu, 2} - \log_{10} E_{\nu, 1}}
\end{equation}
without FSR, where $F_{\nu, 1}$ and $F_{\nu, 2}$ are the neutrino fluxes at $E_{\nu,1}$ and $E_{\nu,2}$, respectively.  However, with FSR, $E_{\nu,1}$ and $E_{\nu,2}$ are shifted by $\delta E_{\nu,1}$ and $\delta E_{\nu,1}$, respectively, and as a result, $F_{\nu, 1}$ and $F_{\nu, 2}$ are shifted by $\delta F_{\nu, 1} = \Gamma \times \delta E_{\nu,1}$ and $\delta F_{\nu, 1} = \Gamma \times \delta E_{\nu,2}$, respectively. 

For spectral-shape measurements using only charged leptons (e.g., throughgoing muons~\cite{IceCube:2021uhz} or Earth-emergent tau leptons), the measured spectral index becomes \begin{equation}
-\Gamma' 
= 
-\Gamma - \frac{1}{\ln 10} \frac{ \Gamma \delta E_{\nu, 2} - \Gamma \delta E_{\nu, 1}}{\log_{10} E_{\nu, 2} - \log_{10} E_{\nu, 1}} \simeq -\Gamma - \frac{0.0075 \Gamma}{\ln10},
\end{equation}
which means that $-\Gamma > -\Gamma'$. Therefore, not including FSR will result in a measured spectral index ($-\Gamma'$) that is softer than the true spectrum ($-\Gamma$). 
Since $\Gamma \sim 2$--3, the distortion is $\sim 0.01$, which is much smaller than the current measurement uncertainty (e.g., IceCube measured $\Gamma = -2.37^{+0.09}_{-0.09}$ using throughgoing muon tracks~\cite{IceCube:2021uhz}). 
Note that the above estimate does not account for the smearing from the neutrino energy to the muon deposit energy, which needs a detailed detector-level simulation.

\section{Collider neutrinos \label{sec:FPF}}
The impact of FSR is detector and observable dependent, depending on energy thresholds and analysis choices. The detectors at the LHC's FPF are very different from the HE and UHE neutrino telescopes discussed above. Therefore, although the FPF will be exposed to neutrinos in the TeV range, similar to the kinematics considered above, their treatment warrants a separate discussion. 

The FPF currently has two operational neutrino detectors (FASER$\nu$ \cite{FASER:2020gpr, FASER:2021mtu} and SND \cite{SNDLHC:2022ihg, SNDLHC:2023pun}), and one proposed but currently unfunded module (FLArE \cite{Batell:2021blf}) that is relevant to the current discussion; successor experiments to both FASER and SND have been proposed \cite{Feng:2022inv}. FASER$\nu$ expects on the order of $2 \times 10^4$ CC-$\mu$ events \cite{Kling:2021gos, FASER:2024ykc}, which is sufficiently large to consider a differential measurement of $\dd \sigma/\dd y$ in analogy with IceCube. A similar event rate is expected at SND \cite{Kling:2021gos}, and FLArE having a larger fiducial volume will obtain roughly ten times the statistics of FASER$\nu$ at fixed beam energy. Both FASER$\nu$ and SND employ nuclear emulsion detectors \cite{FASER:2020gpr, SNDLHC:2022ihg}, whereas the FLArE proposal involves a liquid argon time projection chamber \cite{Batell:2021blf}.

Nuclear emulsion detectors have sub-micron vertex resolution, however photons only become visible in the detector after converting to $e^+e^-$ pairs. This takes place, on average, within roughly one radiation length. The majority, if not all, of CCDIS events involve neutral pions and etas in the final state that decay promptly to two photons. It is therefore nearly impossible to distinguish an FSR photon from photons produced via $\pi^0,\eta\rightarrow \gamma\gamma$. Therefore, despite their impressive vertex resolution, FSR is still best thought of as increasing the apparent energy of a hadronic cascade. 

Nuclear emulsion detectors can, however, isolate the primary $e^\pm$ track from a vertex, since its ionization track is connected directly to the interaction point. This enables these detectors to construct exclusive observables, such as $\dd \sigma / \dd E_e$ in much the same way as neutrino telescopes construct $\dd \sigma/\dd E_\mu$ and $\dd \sigma/\dd E_\tau$. In FLArE, only muon tracks will be distinguishable, and the analysis presented above applies to IceCube. 

At the neutrino energies of interest for the FPF, i.e.~$0.1-2$ TeV, radiative corrections for muons are of similar size to those discussed for neutrino telescopes above. If electron tracks can be reliably separated from the cascade of hadronic and electromagnetic energy deposition, then radiative corrections to electron neutrino inelasticity distributions will be large. The collinear logarithm will be enhanced $\log(s/m_\mu^2)\rightarrow \log(s/m_e^2)$. For $s\sim 2\times m_N\times(1~{\rm TeV})$ using the Sudakov form factor, \cref{eq_sudakov}, as an estimate, one finds that roughly $25\%$ of all events will contain prompt radiation. This suggests a resummation of large logarithms may be necessary. This can be implemented straightforwardly, either analytically or numerically, using existing parton shower codes. Radiative corrections to tau lepton events will be smaller than for muon neutrinos due to the heavier $\tau$ mass and should be amenable to a fixed-order treatment. Some of these issues have already been discussed \cite{Diener:2003ss, Arbuzov:2004zr} in the context of NuTeV's and NOMAD's neutrino DIS measurements.

Let us consider the distortion of conventional DIS variables due to FSR. Hadronic structure functions for nucleons with mass $m_N$ 
are typically parametrized in terms of $Q^2$ and $x$. At the FPF, these variables can be reconstructed in terms of lab-frame quantities as
\begin{align}
    Q^2&=4 E_\nu E_\ell \sin^2\qty(\frac{\theta_\ell}{2})~,\\
    x&=\frac{Q^2}{2m_N E_h}~,
\end{align}
with $E_h$ and experimental ``hadron energy'' and $\theta_\ell$ measured relative to the beam axis. The neutrino energy is estimated as $E_\nu=E_h+E_\ell$, and this quantity is {\it not} influenced by FSR since total energy is conserved.  The true QCD structure functions depend on $Q^2$ and $x$ reconstructed in terms of $E_X$ rather than $E_h$. As already emphasized above in the context of neutrino telescopes, these quantities differ because of FSR from the lepton line that is captured in the hadronic shower, i.e., $E_h = E_\gamma + E_X$. Since most FSR is collinear with the lepton we will neglect changes to $\theta_\ell$ in the following discussion. Treating $E_\gamma$ as a small parameter and holding $E_\nu$ fixed, we then find the following shifts due to FSR, 
\begin{align}
    \qty[\Delta Q^2]_{\rm FSR} &= -4 E_\nu E_\gamma \sin^2\qty(\frac{\theta_\ell}{2})~,\\
    \qty[\Delta x]_{\rm FSR}   &= \frac{\Delta Q^2_{\rm FSR}}{2m_N E_X}  - \frac{E_\gamma}{E_X} x^{(0)}~, 
\end{align}
where $x^{(0)}$ is the value of Bjorken $x$ in the absence of FSR. We therefore see that FSR lowers both $Q^2$ and $x$ in a correlated fashion. The correction to $Q^2$ is comparable to the relative shifts in $y_{\rm avg}$ discussed above, i.e., on the order of a few percent. The shifts to $x$ can be much larger when $E_X \ll E_\nu$ because the highly energetic lepton can radiate photons with energies comparable to $E_X$; in this scenario the equations should not be linearized.

In the absence of FSR, we may re-write the hadronic energy as $E_X=Q^2/(2m_N x)$.  Therefore for $x\ll 1$ we have $E_X \ll E_\nu$. The data at the FPF will roughly cover a region in the $(x,Q^2)$ plane bounded by $ 2\times 10^{-3} \lesssim x \lesssim 0.7$ and $Q^2\lesssim x \times (100~{\rm GeV}^2)$ and typical values are $x \sim 0.1$ and $Q^2 \sim 10-100~{\rm GeV}^2$.  The neutrino beam at the FPF ranges from $100~{\rm GeV}\lesssim E_\nu \lesssim 3~{\rm TeV}$ and peaks around $1~{\rm TeV}$. As an example, taking $E_\nu=1~{\rm TeV}$, $Q^2=30~{\rm GeV}^2$, and $x=0.1$, we find $E_X\simeq 150~{\rm GeV}$ vs.\ a lepton energy of $850~{\rm GeV}$. Therefore even for $E_\gamma = 50~{\rm GeV}$ we can obtain a $\sim 30\%$ distortion in the value of Bjorken $x$.

This suggests that PDF extractions at the FPF will be sensitive to FSR from leptons, and that this effect should be included in future analysis. This is especially interesting in the case of the quark singlet, and total strangeness PDFs where the FPF is expected to contribute world-leading constraints on the underlying PDFs \cite{Cruz-Martinez:2023sdv}. The ultimate impact of FSR on PDF extractions at the FPF warrants a separate dedicated study.

It is also interesting to note that existing neutrino input, specifically dimuon data from  CCFR, NuTeV, and NOMAD \cite{osti_879078, NuTeV:2001dfo, NOMAD:2013hbk}, into both the NNPDF and CTEQ extractions of PDFs  \cite{Hou:2019efy, NNPDF:2021njg} does not include radiative corrections. Currently, these data have only a small impact on the PDF extractions, but if the FPF positions itself as the leading probe of strange quark PDFs \cite{Cruz-Martinez:2023sdv} then radiative corrections will be essential going forward. 
\section{Conclusions and Discussion \label{sec:conclusions} }
High-energy and ultrahigh-energy neutrinos offer valuable insights into cosmic phenomena and fundamental physics, advancing the frontiers of neutrino astrophysics, multimessenger astronomy, and particle physics~\cite{Ackermann:2022rqc, Ackermann:2019ows}. The increasing number of HE and UHE telescopes~\cite{Ackermann:2022rqc} and LHC's FPF~\cite{Feng:2022inv} offer an expanding dataset on HE and UHE neutrinos, which is steadily lowering the floor set by experimental uncertainty. Understanding neutrino interaction and detection is crucial for any neutrino experiment, and increasingly precise cross sections (both total and differential) are required to meet the demands of these ongoing and upcoming HE and UHE neutrino experiments. 

In this paper, we have studied final state radiation (photons) produced by high-energy charged leptons from neutrino CC interactions. When neutrino energies are large compared to lepton masses, $E_\nu \gg m_\ell$, large logarithms appear, which can substantially enhance the probability of prompt internal bremsstrahlung. For example, the QED corrections we discussed here are roughly five times as large as, e.g., the percent-level corrections due to higher-order QCD corrections considered in \cite{Gauld:2019pgt}.
In HE and UHE neutrino telescopes and at the FPF, most of the events are not inclusive (i.e., the charged lepton and/or the rest cascade are separately measurable) so they are impacted by FSR double-logarithms.
We find that the FSR on average shifts the charged-lepton energy by as much as 5\% and the rest cascade energy by as much as 25\%(!) as shown in \cref{Fig_deltaE_Enu} and that FSR is a relevant effect for current experiments given, e.g., IceCube's existing statistical samples and energy resolution.

For HE neutrino observations, FSR shifts the inelasticity measurements, reduces the energy of throughgoing muons, and affects the energy ratio of the two bangs (and decreases the detectability) of the tau-neutrino-induced double-bang signature.
For the inelasticity measurements~\cite{IceCube:2018pgc}, we find that the effect of FSR on the average inelasticity ($y_{\rm avg}$) can be as large as 10\% at energies relevant for IceCube (\cref{Fig_DeltaYavg_Enu}).
We have also investigated the impact of FSR on a measurement of the $\nu/\bar{\nu}$ flux ratio. Using a theory-based analysis (i.e., without accounting for detector resolution or systematic uncertainties), we find that the shift in $y_{\rm avg}$ from FSR could affect the measurements of the HE $\nu/\bar{\nu}$ flux ratio by about 5\%. It would be important to pursue this further with realistic experimental inputs.

For UHE neutrino observations, 
FSR increases the overall detectable energy of the in-ice radio detectors by as much as 20\%, which also effectively lowers the detection threshold.
FSR also affects the flavor measurements (e.g., Refs.~\cite{ANITA:2008mzi, PUEO:2020bnn, RNO-G:2020rmc, IceCube-Gen2:2020qha}), which can measure the charged leptons and hadronic cascade separately with the new proposed method~\cite{Coleman:2024scd}. For detectors that measure the Earth-emergent charged (mainly tau) leptons (e.g., Refs.~\cite{PUEO:2020bnn, POEMMA:2020ykm, Romero-Wolf:2020pzh, GRAND:2018iaj}), FSR lowers the energy of the leptons by as much as 5\%. If unaccounted for, this introduces a systematic bias in the estimate of parent neutrino energies, and should be included to have a more accurate inference of the parent neutrino energy.
Finally, FSR also affects $\nu_\tau$ regeneration~\cite{Halzen:1998be, Bigas:2008sw, Alvarez-Muniz:2018owm} and can result in large energy losses when there are multiple tau interactions within the Earth.

The effects of FSR discussed above will significantly impact the measurements of neutrino fluxes and spectra (Sec.~\ref{sec_SED}). For the former, not accounting for FSR would lead to an underestimation of the neutrino flux by up to 60\%. For the latter, not including FSR will result in a measured spectrum that is softer than the true spectrum when using only charged leptons.
 
In the context of the LHC's FPF, we have discussed the role of FSR when extracting PDFs from neutrino CCDIS data. The most important effect is kinematic, with migration between bins of $x$ and $Q^2$ due to FSR. This effect is well-known and well-studied in the context of electron DIS \cite{Kwiatkowski:1990es}, and that subject's literature can be straightforwardly applied to extractions of PDFs at the FPF. 

Since 2013, IceCube has opened the field of HE neutrino astronomy~\cite{IceCube:2013low}. In 2021, FASER$\nu$ reported the first neutrino candidates from the LHC \cite{FASER:2021mtu}, with subsequent definitive measurements from both the FASER and SND Collaborations in 2023~\cite{FASER:2023zcr, SNDLHC:2023pun}. Soon, UHE neutrinos will also be discovered and then precisely measured. The effects of FSR are large enough to meaningfully impact near-term HE and UHE neutrino analyses as well as measurements at the LHC's FPF, and it will be important to include FSR in analyses of these experiments' data going forward. 

\vspace{0.6cm}
\section*{Acknowledgement}
\vspace{-0.25cm}
The realistic inelasticity distributions of neutrino deep-inelastic scattering~\cite{Xie:2023suk} in this paper were provided to us by Keping Xie, and the atmospheric neutrino flux data were provided to us by Qinrui Liu; we thank both of them for their help.
We are very grateful to John Beacom for discussions, encouragement, and critical feedback on early versions of this work. We thank Kevin McFarland for useful comments about hadronic cascades in nuclear emulsion detectors. We thank Brian Batell, Mauricio Bustamante, John Campbell, Richard Hill, Keith McBride, Lu Lu, Michele Papucci, Mary Hall Reno, Alfonso Garcia-Soto, Christoph Welling, and Tianlu Yuan for helpful discussions. We thank Mauricio Bustamante, Patrick J.~Fox, Francis Halzen, and Spencer Klein for their feedback and comments on the manuscript. 
R.P. is supported by the Neutrino Theory Network under Award Number DEAC02-07CH11359, the U.S. Department of Energy, Office of Science, Office of High Energy Physics, under Award Number DE-SC0011632, and by the Walter Burke Institute for Theoretical Physics.
B.Z. is supported by Fermilab, which is managed by the Fermi Research Alliance, LLC, acting under Contract No.\ DE-AC02-07CH11359.

\textbf{Note added:} After this manuscript was posted on arXiv and submitted to the journal, Ref.~\cite{Weigel:2024gzh} appeared, which includes FSR into CCDIS calculations, and  Ref.~\cite{FerrarioRavasio:2024kem} appeared, which includes FSR directly in a neutrino event generator. These can help facilitate the inclusion of FSR in experimental simulations for many of the phenomenological scenarios discussed in this paper.

\bibliography{refs.bib}

\end{document}